# Automated Human Claustrum Segmentation using Deep Learning Technologies

Ahmed Awad Albishri[1,2], Syed Jawad Hussain Shah[1], Anthony Schmiedler[3], Seung Suk Kang[3], Yugyung Lee[1]
[1]School of Computing and Engineering University of Missouri-Kansas City, USA
[2] College of Computing and Informatics, Saudi Electronic University, Riyadh, Saudi Arabia
[3] Department of Psychiatry Biomedical Sciences, School of Medicine, University of Missouri-Kansas City
{aa8w2,@mail.umkc.edu, a.albishri@seu.edu.sa}, shs6g7@mail.umkc.edu,
anthonyschmiedeler@gmail.com, kangseung@umkc.edu, leeyu@umkc.edu

*Abstract*—In recent years, Deep Learning (DL) has shown promising results in conducting AI tasks such as computer vision and image segmentation. Specifically, Convolutional Neural Network (CNN) models in DL have been applied to prevention, detection, and diagnosis in predictive medicine. Image segmentation plays a significant role in disease detection and prevention. However, there are enormous challenges in performing DL-based automatic segmentation due to the nature of medical images such as heterogeneous modalities and formats, insufficient labeled training data, and the high-class imbalance in the labeled data. Furthermore, automating segmentation of medical images, like magnetic resonance images (MRI), becomes a challenging task. The need for automated segmentation or annotation is what motivates our work. In this paper, we propose a fully automated approach that aims to segment the human claustrum for analytical purposes. We applied a U-Net CNN model to segment the claustrum (Cl) from a MRI dataset. With this approach, we have achieved an average Dice per case score of 0.72 for Cl segmentation, with K=5 for cross-validation. The expert in the medical domain also evaluates these results.

## I. INTRODUCTION

The rapid growth of medical imaging has increased the need for employing advancements in computer vision [1] to improve and ease the image segmentation processes. Current work in this field shows outstanding results using CNNs, from cancer detection [2] to brain tumor segmentation [3] and many other medical problems. However, some objects of medical imaging are challenging for deep learning segmentation approaches, especially those of neuroimaging for research that often aims to identify and delineate small regions of normal brains rather than large lesions or tumors in abnormal brains. Some automatic segmentation or parcellation tools are available for large cortical and subcortical regions (e.g., FreeSurfer [4]), but the current segmentation and parcellation algorithms based on conventional machine learning have limited accuracy. Furthermore, the current automatic parcellation schemes do not involve many small but potentially important brain structures. Hence, to facilitate research and medical practices of such small brain regions, it is imperative to develop an automated method that employs the use of powerful algorithms like CNNs.

Medical images are stored in different formats, as they consist of more than one image or slice, and represent the anatomical volume acquired from imaging machines [5]. Popular medical image formats like JPEG, PNG, TIF, and others represent the image in 2-dimensional arrays [1]. The imaging technologies have provided tremendous knowledge of healthy and diseased anatomies for both research and treatment purposes [1].

Segmentation of images in computer vision is the process of partitioning the image into a set of pixels based on the pixels' similarities, defining the object boundaries in the image. This is achieved by assigning a label or class to every pixel [1]. The sets of pixels represent objects or a boundary of the objects in the image.

In computer vision, Convolutional Neural Networks (CNNs) are driving advances in image recognition [1]. They also dominate in detecting, segmenting, and recognizing the objects within images, with performance comparable to that of humans [1]. In all of these tasks, labeled data are used to teach the machine in a supervised learning manner. Segmentation tasks play a vital role in delineating different anatomical structures and other regions [1]. However, segmentation becomes a challenging task due to the following reasons:

- Variations in the size and structure of the organs
- Different modalities and formats for images
- Insufficient labeled data
- Class imbalance in labels
- Requires domain expertise to segment the data and validate the results manually
- Manual segmentation is a tedious and time-consuming process, being vulnerable to human errors [5]

The claustrum (Cl) is a very good example of such an important target of CNN approach. Cl is a thin deep brain grey matter structure located at the center of each hemisphere. It is known as the brain's most highly connected hub [6]. Cl has reciprocal connectivity with almost all cortical and subcortical brain areas and massive input from all significant neuromodulator circuits [7]. Based on the anatomy and animal neurophysiological findings, Cl has been hypothesized as a brain network hub node for multisensory integration [8], conscious percepts [9], and bottom-up and top-down attention [10].

Cl is a bilateral anatomical structure in the brain that can be identified using magnetic resonance imaging (MRI), as shown in Figure 1. The radiologist uses MRI techniques to detect and diagnose diseases [11]. Although the precise functions of Cl are mostly unknown, numerous studies have reported the significance of Cl in the pathophysiology of various neuropsychiatric disorders. Cl appears to have a critical role in spreading convulsive epileptic seizures. Also, it was reported that electrical stimulation of Cl reversibly disrupted consciousness in a patient with epilepsy [12]. Disruptions in Cl might cause altered consciousness or loss of memory during an epileptic seizure.

Recent studies of post-mortem brains [13], and MRIs [14] found substantial volume reductions of Cl in children with autism spectrum disorders, the neurodevelopmental disorder that is characterized by severe social/cognitive deficits. Also, similar Cl volume deficits were observed in post-mortem brains of people with schizophrenia, especially in those who were diagnosed with paranoid type schizophrenia that has characteristic symptoms of psychosis (e.g., hallucinations and delusions) [15]. These findings suggest that dysfunctional Cl might underlie altered sensory experiences, hallucinations, and severe developmental problems in cognitive and social functions.

The Cl ROI-based approach is to segment the Claustrum using structural MR images manually. An earlier study developed a manual tracing protocol for a volumetric study of the human claustrum [14]. However, the protocol did not provide enough details of the unique structure of Cl and no clear boundary to delineate the sub-regions of Cl.

Despite accumulating evidence of Cl abnormalities in people with neuropsychiatric disorders [16], [17], a very limited number of neuroimaging studies have been conducted to investigate the functions of human Cl. It is primarily due to 1) the methodological limitations of conventional neuroimaging techniques to isolate the thin structure of Cl (i.e., the limited spatial resolutions of MRI and other neuroimaging techniques) and 2) lack of basic neuroimaging tools (e.g., no Cl label in most widely used neuroimaging brain atlases, no reliable method to delineate Cl). Region of interest (ROI) approach to Cl requires manual segmentation of Cl using structural MR images. An earlier study developed a manual tracing protocol for a volumetric study of the human Cl [14]. However, the protocol did not provide enough details of the unique anatomy of Cl and no clear boundary to delineate the sub-regions. Also, the manual segmentation of Cl on MRI is a time- consuming and challenging procedure that requires domain expertise.

In our case, pixels were associated with either one of two classes, Cl or background. In this paper, we propose a U-Net CNN architecture to achieve automatic and accurate segmentation of human Cl that is particularly challenging to delineate in MR images due to its thin morphology. To this end, we trained and validated our U-Net DL model using the Cl label maps manually segmented based on a manual tracing protocol. To improve the model accuracy with the limited number of the training dataset, we optimized the MRI preprocessing steps, including data augmentation strategy. The model performance was evaluated using quantitative similarity indices comparing the model generated with the ground truth input data.

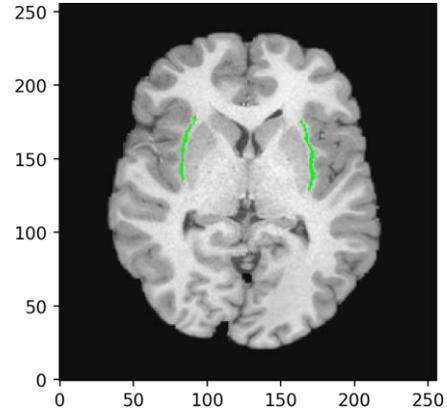

Fig. 1: The claustrum delineated in a T1-weighted MRI (highlighted in green).

## II. RELATED WORK

Many of the recent work has used CNN based architecture for image segmentation. In this section, we will discuss some of such approaches. Fully Convolutional Neural Networks (FCNs) by Long et al. [18], was the first successful model in deep learning image segmentation, which introduced the idea of skip connection to up-sample the network output. The main drawback of this architecture includes the difficulty in training the model from scratch. Similarly, Segnet [19] incorporated the idea of skip connections with a novel implementation that copied the indices from max pooling to the right side (decoder) to produce a better segmentation and make the network faster and lighter. U-Net architecture by Ronneberger et al. [20] employed the idea of skip connection by copying the feature maps from encoder to decoder part. This model was designed to work successfully with a minimal amount of biomedical data. Data augmentation was the key to the success of this model.

Moeskops et al. [21] applied a CNN model that used multiple patch sizes and various convolution kernel sizes to get multi-scale information of each voxel. The authors successfully created a robust model that worked with different image modality and achieved a dice score of 80% in various datasets. Kayalibay et al. [22] applied the U-Net model to segment two datasets, including hand and brain scans. They made two changes to U-Net architecture, which included combining multiple feature maps at different scales and element-wise multiplication instead of concatenation. The first change harmed the model performance, while the second one speeded up the convergence time. Kushibar et al. [23] introduced a novel method that took advantage of combining the convolutional features and the prior spatial features from a brain atlas. Their network was trained with 2.5D batches, instead of 3D, due to the memory constraint. Christ et al. [24] applied the U-Net model to segment the liver and liver lesions in a cascaded way, and they reported improved accuracy using a conditional random field (CRF). Recently, we presented a novel cascaded U-Net model for liver and liver tumor segmentation and summarization [25].

### III. APPROACHES

#### A. Data

We used T1-weighted MRIs of 30 healthy adults (13 males and 17 females; age range 21-35 years old) to develop the automatic segmentation model of Cl. The dataset was collected as parts of the Washington University-Minnesota Consortium Human Connectome Project (WU-Minn HCP) [26]. We downloaded the randomly selected 30 subjects' T1 MRIs from the ConnectomDB [27]. The MRIs were acquired using a customized Siemens 3T Connectome Skyra scanner with the 3D MPRAGE T1-weighted sequence with 0.7 mm isotropic resolution (FOV=224 mm, matrix=320, 256 sagittal slices in a single slab, TR=2400 ms, TE=2.14 ms, TI=1000 ms, flip angle=8°). The details of MRI data collection and preprocessing methods are described in [28].

In [29], we manually segmented Cl of the 30 T1 MRI scans based on a protocol developed for the conventional 3T resolution (1.0 mm isotropic voxels) T1 MRIs based on a conventional brain atlas [30] widely used in the field. We used the annotated Cl label map in Nifti 3D format to train the segmentation model.

#### B. Preprocessing

We preprocessed the data in a slice-wise fashion for the optimal model performance. The original dataset came in two sizes, 218×364 and 311×260. Initially, all the input slices were resized to 256×256 resolution to have a consistent image resolution for the whole dataset. The labels were assigned a binary value of 0 and 1 for background and Cl, respectively. Normalization or feature scaling was performed on pixel intensities. It is an effective popular technique in data preparation for machine learning algorithms. The normalization step only standardizes or rescales the range of data features and in our case, the images' pixel intensities, to the standard and most used range of 0 to 1.

This step is helpful because standardizing all the values of the intensities of the pixels enhances the model performance, optimizes the model weights while training, and removes the irrelevant aspect of the data [31]. Further, data was augmented to teach the model the desired invariance properties and to increase the training dataset. The augmentation process involved elastic deformation, image transformation, and intensity rescaling. In deep learning algorithms, the more data the algorithm trains on, the better the results it can produce [32]. Data augmentation is also a general solution to reduce overfitting on image data [33]. It is often difficult to obtain a large number of accurately labeled medical images (e.g., manually segmented MRI data). Therefore, it is highly recommended to apply the data augmentation techniques for development of an automatic segmentation algorithm of medical imaging data [20].

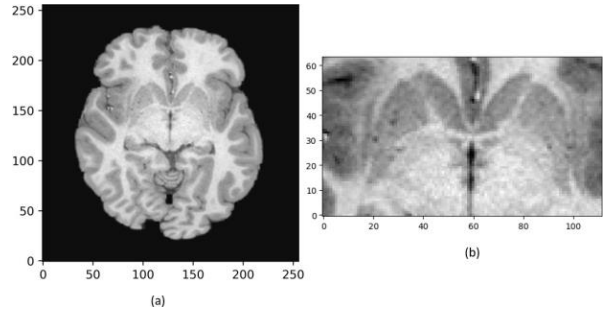

Fig. 2: Comparison between MRI slice. (a) before ROI, size of 256×256, (b) the same slice after ROI, size of 64x112.

We had a massive imbalance in the data between the target (Cl: 0.3%) and the background (the other brain and the black background regions: 99.7%) due to the small size of Cl. To address this class imbalance issue, we have adopted several strategies. First, we selected only the axial plane slices that contain Cl on the images and used them as training data for the model. Second, instead of training the model on the full-size slices (256×256), the input images were cropped to obtain the region of interests (ROI) of size 64×112. After applying the ROI, each slice contained approximately 97% of background pixels and 3% of Cl pixels. As shown in Table 1 comparing the number pixels of Cl and the background before and after the ROI procedure, the ROI procedure greatly reduced the number of background pixels almost by a factor of 10 without affecting the number of Cl pixels, reducing the class imbalance. Figure 2 shows the comparison between a slice before ROI (size 256×256), and the same slice after ROI (64×112). After the model training and validation of the ROI slices, we restored them to their original size of 256×256.

#### C. Model Architecture

Convolution network models dominate the field of computer vision, and every model comes with its architecture and advantages. The architecture of our model is shown in Figure 3. It is a U-Net model, which is composed of an encoder and decoder path with skip connections in between. The encoder path, on the left, is called the contracting path, and it captures the context from the image. The decoder path, on the right, is called the expanding path, and it enables the precise localization of object that needs to be segmented. Features learned at each level in the encoder path are transferred to the decoder path through skip connections, where they will be concatenated with features from the decoder path [20].

As shown in Figure 3, our model is four layers deep. Each layer in the encoder path performs two rounds of convolution, batch normalization, and dropout, respectively. After that, max-pooling reduces the image size by half and passes it to the next layer in the architecture. This process continues until the image reaches the last layer in the encoder path, called the bottleneck layer. From this point on in the model, the decoder path starts. Initially, upsampling is applied to increase the image resolution by a factor of 2. In regular upsampling, convolutions are replaced by transposed convolution to up-sample the image. After that, the image is transferred to the next layer in the decoder path, which performs two rounds of convolution, batch normalization, and dropout, respectively. This process continues until the image reaches the last level in the decoder path, where it is restored to its original size. The last layer in the decoder path has a sigmoid function that classifies each pixel in a binary way to give it a probability between 0 to 1.

TABLE I: The number of Cl and the background pixels before and after an ROI procedure for 3 slices. The number of clastrum pixels did not change before and after ROI but the number of background pixels are reduced by almost a factor of 10 after ROI.

| Slice ID | Number of Cl Pixels Before ROI | Number of Cl Pixels After ROI | Number of the Background Pixels Before ROI | Number of the Background Pixels After ROI |
|---|---|---|---|---|
| 1 | 198 | 198 | 65338 | 6970 |
| 2 | 209 | 209 | 65327 | 6959 |
| 3 | 187 | 187 | 65349 | 6981 |

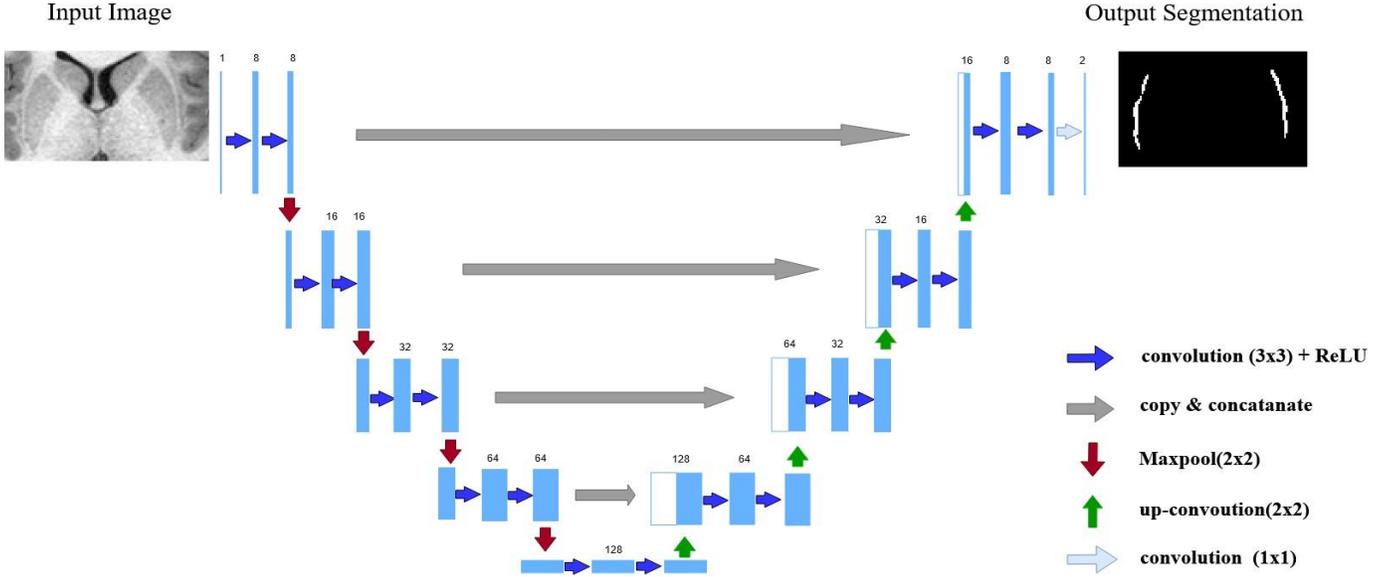

Fig. 3: Model architecture consisting of a contracting path (left) and expanding path (right) with skip connections in between (middle). The model takes an MRI as an input and outputs the segmentation of Cl.

In the convolution process, we applied a window of size 3×3 to the input image to construct a feature map. We used the ReLU activation function [34] in these convolution processes. This activation function does not change the values of the positive weights of the features but will assign a value of zero to all the negative weights of the features. Batch normalization is used to accelerate the network training by reducing the rate at which each layer's inputs distribution changes during training, as the previous layer's parameters change [35].

To prevent overfitting, we have implemented L2 and Dropout as regularization methods in the model. We have also used early stopping [36] for the same reason. We have used Adam optimizer to optimize the weight and bias at each layer in the network, with a learning rate of 0.001 [36]. The input to the model is brain MRI and output is the segmented Cl image.

This architecture enables the model to produce good segmentation results with a small amount of training data. The model performs inter- and intra-slice classification to detect Cl. Also, it is trained to identify the coarse Cl boundaries and segment it from 2.5D images from the dataset. It takes a NiftTi format T1 MRI file as an input for a subject and classifies it to estimate the probability of each pixel belongs to a Cl or background at each slice in the output of the subject.

### D. Loss Function

Binary Cross-Entropy (BCE) loss function [37] is used to enhance model performance for binary classification. The loss function reduces the overall classification error caused by a huge class imbalance between the target (Cl) and the background pixels. BCE is calculated by the following:

$$f(c,\hat{c}) = -(1-w)*c\log\hat{c} - w*(1-c)\log(1-\hat{c}) \quad (1)$$

where $\hat{c}$ represents the weighted term for the model prediction of Cl, and $c$ represents the weighted term for the ground truth. A weight variable $w$ provides a value for each observation in a data set. We computed the weighted terms for the foreground (Cl) and the background classes to train each segmented network. The weighted term of Cl was obtained by summing all Cl pixels and dividing them by a total number of pixels (Cl + the background). It is noteworthy that only the ROI slices that contained Cl were considered, as proposed in [38]. We also calculated the weighted term for the background classes similarly. The weighted terms were then normalized to obtain weight $w$ and $1 - w$ to balance the BCE.

## IV. RESULTS AND DISCUSSION

Our model has accomplished the average dice per case score of 0.72 for Cl segmentation, by implementing K-Fold cross-validation with the value of K set to 5. Using the HCP MRI data, we have achieved competitive Cl segmentation scores. As shown in Figure 4 depicting the ground truth and the model segmentation results of two subjects, the segmentation models precisely assigned each pixel of an image to one of the two classes, Cl and the background.

To quantitatively evaluate the model accuracy, we calculated two accuracy indices, including dice score, and intraclass correlation coefficient (ICC). Dice score is an index used to gauge the similarity of two samples in a range from 0 to 1. The higher the value for dice scores, the better is the accuracy of the model. Equation 2 shows the formula for calculating the dice score. In Equation 2, TP stands for true positive, the number of ground truth Cl pixels correctly predicted by the model. FP stands for false positive, the number of background pixels that the model incorrectly predicted as Cl, while FN stands for false negative, the number of Cl pixels that the model incorrectly predicted as the background.

$$DiceScore = \frac{2TP}{2TP + FP + FN} \quad (2)$$

We calculated dice scores from K-Fold cross-validation performed on the training set of 30 MRIs. For Cl segmentation, we trained the model on the augmented dataset and tested it on a non-augmented one. The value of K was set to 5 for cross-validation, in which the dataset was divided into five equal parts, and the model was trained on four parts and tested on the other one part. The whole process was repeated five times to ensure that the model is tested on the whole dataset. Dice score was calculated at each iteration. With this procedure, we obtained an average dice per case score of 0.72 for the Cl segmentation model. We calculated ICC that measures the reliability of ratings or measurements for classes or clusters (i.e., data collected as groups or sorted into groups), describing how the observations in the same class resemble each other [39]. Like the dice score, ICC ranges from 0 to 1, with 1 being the perfect agreement between classes. Using the Shrout-Fleiss 2K ICC formula [39], we obtained ICC of 0.81 between the model and the ground truth.

These results demonstrated the strengths of our U-Net approach and optimized preprocessing strategies to solve the challenging problem to segment Cl. Compared to other subcortical brain structures that are often segmented manually (e.g., amygdala), Cl is particularly difficult to segment accurately due to the thin morphology and the limited spatial resolutions of MR images. For example, the proportion of Cl MRI voxels forming the boundaries of Cl (74%) is much higher than that of amygdala (29%) whose volume is similar to that of Cl. Given that the segmentation errors (i.e., false positives and false negatives) primarily occur at the boundaries of the target structure, we can anticipate about 2.5 times higher probability to have segmentation errors for Cl than amygdala. Our results of dice score greater than .72 and ICC greater than 0.81 that were obtained from the limited training dataset of 30 subjects, therefore, provide evidence of the great potential of our DL approach for development of automatic and accurate segmentation tools for very challenging medical imaging segmentation problems, including Cl segmentation.

On the other hand, Cl was challenging for machine learning due to its large class imbalance problem (i.e., the much higher proportions of the background pixels compared to Cl). Although our ROI strategy reduced the severity of the problem, future studies may need to use additional advanced DL techniques (e.g., Attention Gate modeling) to improve the results. It is noteworthy that potential human errors in the training dataset might also contribute to the limited DL model accuracy. The training dataset used in the study was suboptimal in that we manually segmented Cl of the high-resolution HCP MR images (0.7 mm isotropic voxels) using the manual tracing protocol developed based on the conventional brain atlas for lower resolution MRI images (1.0 mm isotropic voxels). We recently updated the manual segmentation protocol [29] using a cellular-level human brain atlas [40] that incorporated neuroimaging (T1- and diffusion-weighted MRI), high-resolution histology, large-format cellular resolution Nissl and immunohistochemistry anatomical plates of an intact adult brain. The recently updated Cl manual segmentation protocol has led superior Cl segmentation results compared to the old protocol. With an optimal training dataset (i.e., Cl label maps that are more accurately delineated with the updated protocol), we may be able to achieve even higher DL model accuracy.

TABLE II: Intraclass Correlation Coefficients (ICC)

| Type | | ICC |
|---|---|---|
| Single_raters_absolute | ICC1 | 0.65 |
| Single_random_raters | ICC2 | 0.69 |
| Single_fixed_raters | ICC3 | 0.86 |
| Average_raters_absolute | ICC1k | 0.79 |
| Average_random_raters | ICC2k | 0.81 |
| Average_fixed_raters | ICC3k | 0.92 |
| Number of subjects = 30 | Number of Judges = 2 | |

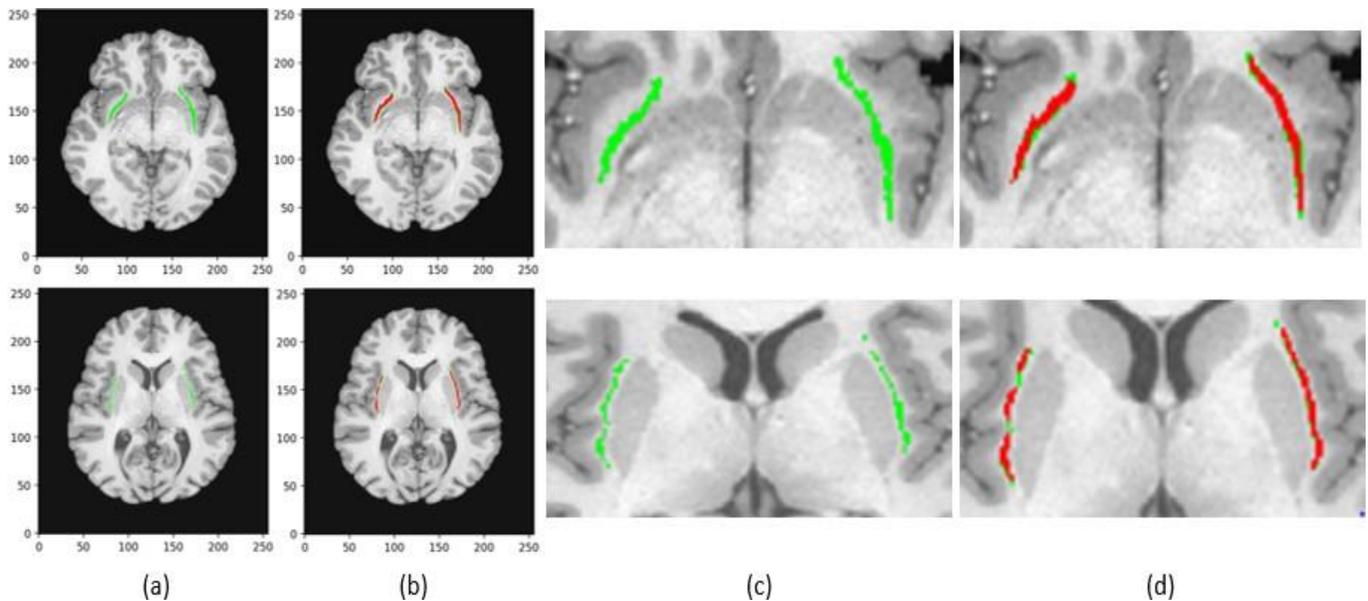

Fig. 4: Example results of the U-Net automatic segmentation of Cl from two participants. From left to right: Column (a) shows the ground truth in green, and Column (b) shows the model prediction in red on top of the ground truth. Column (c) shows the ROI of (a), and Column (d) shows the ROI of (b). Each row represents one particular slice of a subject.

## V. CONCLUSION

We proposed a 2-dimensional network that consists of U-Net architecture to perform the segmentation of Cl. The model takes the brain MRI label maps of Cl as input and outputs Cl segmentation. Our model achieved promising results in terms of dice score and ICC for segmentation tasks for Cl, the particularly difficult subcortical structure for a manual and an automatic segmentation. Also we suffered problems related to the class imbalance between Cl and the background. We have implemented ROI to tackle the imbalance issue to some extent. It is recommended for future studies to utilize additional advanced DL techniques as well as our optimized procedures for challenging medical image segmentation problems.